\documentclass[aps,pra,showpacs,superscriptaddress,preprint]{revtex4}
\usepackage{amssymb}
\usepackage{amsmath}
\usepackage{graphicx}

\catcode`ð=\active
 \defð{\u{g}}
 \catcode`Ð=\active
 \defÐ{\u{G}}
 \catcode`Ý=\active
\defÝ{\. I}
 \catcode`ö=\active
\defö{\"{o}}
 \catcode`Ö=\active
 \defÖ{\"O}
 \catcode`ü=\active
 \defü{\"{u}}
 \catcode`Ü=\active
 \defÜ{\"{U}}
 \catcode`Þ=\active
\defÞ{\c{S}}
 \catcode`þ=\active
 \defþ{\c{s}}
 \catcode`ý=\active
 \defý{{\i}}
 \catcode`ç=\active
\defç{\d{c}}
 \catcode`Ç=\active
\defÇ{\d{C}}

\input{tcilatex}

\begin{document}

\title{Bound states of the Klein-Gordon equation in $D$-dimensions with some
physical scalar and vector exponential-type potentials including orbital
centrifugal term }
\author{Sameer M. Ikhdair}
\email[E-mail: ]{sikhdair@neu.edu.tr}
\affiliation{Physics Department, Near East University, Nicosia, North Cyprus, Turkey}
\date{%
\today%
}

\begin{abstract}
The approximate analytic bound state solutions of the Klein-Gordon equation
with equal scalar and vector exponential-type potentials including the
centrifugal potential term are obtained for any arbitrary orbital angular
momentum number $l$ and dimensional space $D.$ The
relativistic/non-relativistic energy spectrum equation and the corresponding
unnormalized radial wave functions, in terms of the Jacobi polynomials $%
P_{n}^{(\alpha ,\beta )}(z),$ where $\alpha >-1,$ $\beta >-1$ and $z\in
\lbrack -1,+1]$ or the generalized hypergeometric functions $%
_{2}F_{1}(a,b;c;z),$ are found. The Nikiforov-Uvarov (NU) method is used in
the solution$.$ The solutions of the Eckart, Rosen-Morse, Hulth\'{e}n and
Woods-Saxon potential models can be easily obtained from these solutions.
Our results are identical with those ones appearing in the literature.
Finally, under the PT-symmetry, we can easily obtain the bound state
solutions of the trigonometric Rosen-Morse potential.

\end{abstract}

\maketitle

\newpage

\section{Introduction}

The exact solutions of the wave equations (non-relativistic or relativistic)
are very important since they contain all the necessary information
regarding the quantum system under consideration. However, analytical
solutions are possible only in a few simple cases such as the hydrogen atom
and the harmonic oscillator [1,2]. Most quantum systems could be solved only
by using approximation schemes like rotating Morse potential via Pekeris
approximation [3] and the generalized Morse potential by means of an
improved approximation scheme [4]. \ Recently, the study of exponential-type
potentials have attracted much attention from many authors [5-26]. These
potentials include the Woods-Saxon [5,6], Hulth\'{e}n [7-16], Manning-Rosen
[17-22], the Eckart [23-25] and the Rosen-Morse [26] potentials.

The spherically symmetric Eckart-type potential model [27] is a molecular
potential model which has been widely applied in physics [28] and chemical
physics [29,30] and is generally expressed as%
\begin{equation}
V(r;q)=V_{1}\cos ech_{q}^{2}\alpha r-V_{2}\coth _{q}\alpha r,\text{ }%
V_{1},V_{2}>0,\text{ }-1\leq q<0\text{ or }q>0,
\end{equation}%
where the coupling parameters $V_{1}$ and $V_{2}$ describe the depth of the
potential well, while the screening parameter $\alpha $ is related to the
range of the potential. It is a special case of the five-parameter
exponential-type potential model [31,32]. The range of parameter $q$ was
taken as $q>0$ by Ref. [33] and has been extended to $-1\leq q<0$ or $q>0$\
or even complex by Ref. [34]. The deformed hyperbolic functions given in (1)
have been introduced for the first time by Arai [35] for real $q$ values.
When\ $q$ is complex, the functions in (1) are called the generalized
deformed hyperbolic functions. The Eckart-type potentials (1) can also be
written in the exponential form as%
\begin{equation}
V(r;q)=4V_{1}\frac{e^{-2\alpha r}}{\left( 1-qe^{-2\alpha r}\right) ^{2}}%
-V_{2}\frac{1+qe^{-2\alpha r}}{1-qe^{-2\alpha r}}.
\end{equation}%
The study of the bound and scattering states for the Eckart-type potential
has raised a great deal of interest in the non-relativistic as well as in
relativistic quantum mechanics. The $s$-wave ($l=0$) bound-state solution of
the Schr\"{o}dinger equation for the Eckart potential has been widely
investigated by using various methods, such as the supersymmetric (SUSY)
shape invariance technology [36], point cannonical transformation (PCT)
method [37] and SUSY Wentzel-Kramers-Brillouin (WKB) approximation approach
[38]. The bound state solutions of the $s$-wave Klein-Gordon (KG) equation
with equally mixed Rosen-Morse-type (Eckart and Rosen--Morse well)
potentials have been studied [39]. The bound state solutions of the $s$-wave
Dirac equation with equal vector and scalar Eckart-type potentials in terms
of the basic concepts of the shape-invariance approach in the SUSYQM have
also been studied [24]. The spin symmetry and pseudospin symmetry in the
relativistic Eckart potential have been investigated by solving the Dirac
equation for mixed potentials [25]. Unfortunately, the wave equations for
the Eckart-type potential can only be solved analytically for zero angular
momentum states because of the centrifugal potential term. Some authors
[23-25] studied the analytical approximations to the bound state solutions
of the Schr\"{o}dinger equation with Eckart potential by using the usual
existing approximation scheme proposed by Greene and Aldrich [40] for the
centrifugal potential term. This approximation has also been used to study
analytically the arbitrary $l$-wave scattering state solutions of the Schr%
\"{o}dinger equation for the Eckart potential [41,42]. The same
approximation scheme for the spin-orbit coupling term has been used to study
the spin symmetry and pseudospin symmetry analytical solutions of the Dirac
equation with the Eckart potential using the AIM [43]. Overmore, the
pseudospin symmetry analytical solutions of the Dirac equation for the
Eckart potential have been found by using the SUSY WKB formalism [44]. Very
recently, for the first time, the approximation scheme for the centrifugal
potential term has also been used in [45] to obtain the approximate
analytical solution of the KG equation for equal scalar and vector Eckart
potentials for arbitrary $l$-states by means of the functional analysis
method.

This approximation for the centrifugal potential term [7,14,40] has also
been used to solve the Schr\"{o}dinger equation [7,14], KG [8,15] and Dirac
equation [15] for the Hulth\'{e}n potential. Recently, the KG and Dirac
equations have been solved in the presence of the Hulth\'{e}n potential,
where the energy spectrum and the scattering wave functions were obtained
for spin-$0$ and spin-$\frac{1}{2}$ particles, using a more general
approximation scheme for the centrifugal potential [15]. They found that the
good approximation, however, occurs when the screening parameter $\alpha $
and the dimensionless parameter $\gamma $ are taken as $\alpha =0.1$ and $%
\gamma =$1, respectively, which is simply the case of the usual
approximation [7,14]. Also, other authors have recently proposed an
alternative approximation scheme for the centrifugal potential to solve the
Schr\"{o}dinger equation for the Hulth\'{e}n potential [46]. Taking $\omega
=1,$ their approximation can be reduced to the usual approximation [7,14].
Very recently, we have also proposed a new approximation scheme for the
centrifugal term [9].

The Nikiforov-Uvarov (NU) method [47] and other methods have also been used
to solve the $D$-dimensional Schr\"{o}dinger equation [48] and relativistic $%
D$-dimensional KG equation [49], Dirac equation [4,10,26,50] and spinless
Salpeter equation [51].

The aim of this work is to employ the usual approximation scheme [40,45] in
order to solve the $D$-dimensional radial KG equation\ for any orbital
angular momentum number $l$ for the scalar and vector Eckart-type potentials
using a general mathematical model of the NU method. This offers a simple,
accurate and efficient scheme for the exponential-type potential models in
quantum mechanics. We consider the following relationship between the scalar
and vector potentials: $V(r)=V_{0}+\beta S(r),$ where $V_{0}$ and $\beta $
are arbitrary constants [52]. Under the restriction of equally mixed
potentials $S(r)=V(r),$ the KG equation turns into a Schr\"{o}dinger-like
equation and thus the bound state solutions are very easily obtained through
the well-known methods developed in the non-relativistic quantum mechanics.
It is interesting to note that, this restriction include the case where $%
V(r)=0$ when both constants vanish, the situation where the potentials are
equal $(V_{0}=0;\beta =1)$ and also the case where the potentials are
proportional [53] when $V_{0}=0$ and $\beta =\pm 1,$ which provide the
equally-mixed scalar and vector potential case $V(r)=\pm S(r)$. Very
recently, we have obtained an approximate analytic solution of the KG
equation in the presence of equal scalar and vector generalized deformed
hyperbolic potential functions by means of parameteric generalization of the
NU method. Furthermore, for the equally-mixed scalar and vector potential
case $V(r)=\pm S(r),$ we have obtained the approximate bound state
rotational-vibrational (ro-vibrational) energy levels and the corresponding
normalized wave functions expressed in terms of the Jacobi polynomial $%
P_{n}^{\left( \mu ,\nu \right) }(x),$ where $\mu >-1,$ $\nu >-1$ and $x\in %
\left[ -1,+1\right] $ for a spin-zero particle in a closed form [54].

The paper is structured as follows: In section 2, we derive a general model
of the NU method valid for any central or non-central potential. In section
3, the approximate analytical solutions of the $D$-dimensional radial KG
equation with arbitrary $l$-states for equally-mixed scalar and vector
Eckart-type potentials and other typical potentials are obtained by means of
the NU method. Also, the exact $s$-wave KG equation has also been solved for
the Rosen-Morse-type potentials and other typical potentials. The relative
convenience of the Eckart-type potential (Rosen-Morse-type potential) with
the Hulth\'{e}n potential (Woods-Saxon potential) has been studied,
respectively. We make some remarks on the energy equations and the
corresponding wavefunctions for the Eckart and Rosen-Morse well potentials
in various dimensions and their non-relativistic limits in section 4.
Section 5 contains the summary and conclusions.

\section{NU Method}

The NU method is briefly outlined here and the details can be found in [47].
This method was proposed to solve the second-order differential wave
equation of the hypergeometric-type: 
\begin{equation}
\sigma ^{2}(z)\psi _{n}^{\prime \prime }(z)+\sigma (z)\widetilde{\tau }%
(z)\psi _{n}^{\prime }(z)+\widetilde{\sigma }(z)\psi _{n}(z)=0,
\end{equation}%
where $\sigma (z)$ and $\widetilde{\sigma }(z)$ are at most second-degree
polynomials and $\widetilde{\tau }(z)$ is a first-degree polynomial. The
prime denotes the differentiation with respect to $z.$ To find a particular
solution of Eq. (3), one can decompose the wave function $\psi _{n}(z)$ as
follows:%
\begin{equation}
\psi _{n}(z)=\phi _{n}(z)y_{n}(z),
\end{equation}%
leading to a hypergeometric type equation 
\begin{equation}
\sigma (z)y_{n}^{\prime \prime }(z)+\tau (z)y_{n}^{\prime }(z)+\lambda
y_{n}(z)=0,
\end{equation}%
where%
\begin{equation}
\lambda =k+\pi ^{\prime }(z),
\end{equation}%
and $y_{n}(z)$ satisfies the Rodrigues relation%
\begin{equation}
y_{n}(z)=\frac{A_{n}}{\rho (z)}\frac{d^{n}}{dz^{n}}\left[ \sigma ^{n}(z)\rho
(z)\right] .
\end{equation}%
In the above equation, $A_{n}$ is a constant related to the normalization
and $\rho (z)$ is the weight function satisfying the condition%
\begin{equation}
\sigma (z)\rho ^{\prime }(z)+\left( \sigma ^{\prime }(z)-\tau (z)\right)
\rho (z)=0,
\end{equation}%
with 
\begin{equation}
\tau (z)=\widetilde{\tau }(z)+2\pi (z),\tau ^{\prime }(z)<0.
\end{equation}%
Since $\rho (z)>0$ and $\sigma (z)>0,$ the derivative of $\tau (z)$ should
be negative [47] which is the essential condition for a proper choice of
solution. The other part of the wavefunction in Eq. (4) is defined as%
\begin{equation}
\sigma (z)\phi ^{\prime }(z)-\pi (z)\phi (z)=0,
\end{equation}%
where%
\begin{equation}
\pi (z)=\frac{1}{2}\left[ \sigma ^{\prime }(z)-\widetilde{\tau }(z)\right]
\pm \sqrt{\frac{1}{4}\left[ \sigma ^{\prime }(z)-\widetilde{\tau }(z)\right]
^{2}-\widetilde{\sigma }(z)+k\sigma (z)}.
\end{equation}%
The determination of $k$ is the essential point in the calculation of $\pi
(z),$ for which the discriminant of the square root in the last equation is
set to zero. This results in the polynomial $\pi (z)$ which is dependent on
the transformation function $z(r).$ Also, the parameter $\lambda $ defined
in Eq. (6) takes the following form%
\begin{equation}
\lambda =\lambda _{n}=-n\tau ^{\prime }(z)-\frac{1}{2}n\left( n-1\right)
\sigma ^{\prime \prime }(z),\ \ \ n=0,1,2,\cdots .
\end{equation}%
We may construct a general recipe of the NU method valid for any central and
non-central potential. We begin by comparing the following hypergeometric
equation 
\begin{equation}
\left[ z\left( 1-c_{3}z\right) \right] ^{2}\psi _{n}^{\prime \prime }(z)+%
\left[ z\left( 1-c_{3}z\right) \left( c_{1}-c_{2}z\right) \right] \psi
_{n}^{\prime }(z)+\left( -Az^{2}+Bz-C\right) \psi _{n}(z)=0,
\end{equation}%
with its counterpart Eq. (3), we then obtain [54]%
\begin{equation}
\widetilde{\tau }(z)=c_{1}-c_{2}z,\text{ }\sigma (z)=z\left( 1-c_{3}z\right)
,\text{ }\widetilde{\sigma }(z)=-Az^{2}+Bz-C.
\end{equation}%
Substituting Eq. (14) into Eq. (11), we find%
\begin{equation}
\pi (z)=c_{4}+c_{5}z\pm \left[ \left( c_{6}-c_{3}k_{+,-}\right) z^{2}+\left(
c_{7}+k_{+,-}\right) z+c_{8}\right] ^{1/2},
\end{equation}%
where%
\begin{equation}
c_{4}=\frac{1}{2}\left( 1-c_{1}\right) ,\text{ }c_{5}=\frac{1}{2}\left(
c_{2}-2c_{3}\right) ,\text{ }c_{6}=c_{5}^{2}+A,\text{ }c_{7}=2c_{4}c_{5}-B,%
\text{ }c_{8}=c_{4}^{2}+C.
\end{equation}%
The discriminant under the square root sign must be set to zero and the
resulting equation must be solved for $k,$ it yields%
\begin{equation}
k_{+,-}=-\left( c_{7}+2c_{3}c_{8}\right) \pm 2\sqrt{c_{8}c_{9}},
\end{equation}%
where%
\begin{equation}
c_{9}=c_{3}\left( c_{7}+c_{3}c_{8}\right) +c_{6}.
\end{equation}%
Inserting Eq. (17) into Eq. (15) and solving the resulting equation, we make
the following choice of parameters: 
\begin{equation}
\pi (z)=c_{4}+c_{5}z-\left[ \left( \sqrt{c_{9}}+c_{3}\sqrt{c_{8}}\right) z-%
\sqrt{c_{8}}\right] ,
\end{equation}%
\begin{equation}
k_{-}=-\left( c_{7}+2c_{3}c_{8}\right) -2\sqrt{c_{8}c_{9}}.
\end{equation}%
Further, from Eq. (9), we get 
\begin{equation}
\tau (z)=1-\left( c_{2}-2c_{5}\right) z-2\left[ \left( \sqrt{c_{9}}+c_{3}%
\sqrt{c_{8}}\right) z-\sqrt{c_{8}}\right] ,
\end{equation}%
whose derivative must be negative:%
\begin{equation}
\tau ^{\prime }(z)=-2c_{3}-2\left( \sqrt{c_{9}}+c_{3}\sqrt{c_{8}}\right) <0,
\end{equation}%
in accordance with essential requirement of the method [47]. Solving Eqs.
(6) and (12), we get the energy equation:%
\begin{equation}
\left( c_{2}-c_{3}\right) n+c_{3}n^{2}-\left( 2n+1\right) c_{5}+\left(
2n+1\right) \left( \sqrt{c_{9}}+c_{3}\sqrt{c_{8}}\right) +c_{7}+2c_{3}c_{8}+2%
\sqrt{c_{8}c_{9}}=0,
\end{equation}%
for the potential under investigation. Let us now turn to the wave
functions. The solution of the differential equation (8) for the weight
function $\rho (z)$ is%
\begin{equation}
\rho (z)=z^{c_{10}}(1-c_{3}z)^{c_{11}},
\end{equation}%
and consequently from Eq. (7), the first part of the wave function becomes 
\begin{equation}
y_{n}(z)=P_{n}^{\left( c_{10},c_{11}\right) }(1-2c_{3}z),\text{ }\func{Re}%
(c_{10})>-1,\text{ }\func{Re}(c_{11})>-1,
\end{equation}%
where 
\begin{equation}
c_{10}=c_{1}+2c_{4}+2\sqrt{c_{8}}-1,\text{ }c_{11}=1-c_{1}-2c_{4}+\frac{2}{%
c_{3}}\sqrt{c_{9}},
\end{equation}%
and $P_{n}^{\left( a,b\right) }(1-c_{3}z)$ are Jacobi polynomials. The
second part of the wave function (4) can be found from the solution of the
differential equation (10) as%
\begin{equation}
\phi (z)=z^{c_{12}}(1-c_{3}z)^{c_{13}},
\end{equation}%
where 
\begin{equation}
c_{12}=c_{4}+\sqrt{c_{8}},\text{ }c_{13}=-c_{4}+\frac{1}{c_{3}}\left( \sqrt{%
c_{9}}-c_{5}\right) .
\end{equation}%
Hence, the general wave functions (4) read as%
\begin{equation}
u_{l}(z)=N_{n}z^{c_{12}}(1-c_{3}z)^{c_{13}}P_{n}^{\left(
c_{10},c_{11}\right) }(1-2c_{3}z),
\end{equation}%
where $N_{n}$ is a normalization constant.

\section{Bound-State Solutions}

The $D$-dimensional time-independent arbitrary $l$-states radial KG equation
with scalar and vector potentials $S(r)$ and $V(r),$ respectively, where $%
r=\left\vert \mathbf{r}\right\vert $ describing a spinless particle takes
the general form [3,49]: 
\begin{equation*}
\mathbf{\nabla }_{D}^{2}\psi _{l_{1}\cdots l_{D-2}}^{(l_{D-1}=l)}(\mathbf{x}%
)+\frac{1}{\hbar ^{2}c^{2}}\left\{ \left[ E_{nl}-V(r)\right] ^{2}-\left[
Mc^{2}+S(r)\right] ^{2}\right\} \psi _{l_{1}\cdots l_{D-2}}^{(l_{D-1}=l)}(%
\mathbf{x})=0,
\end{equation*}%
\begin{equation}
\text{ }\nabla _{D}^{2}=\sum\limits_{j=1}^{D}\frac{\partial ^{2}}{\partial
x_{j}^{2}},\text{ }\psi _{l_{1}\cdots l_{D-2}}^{(l_{D-1}=l)}(\mathbf{x}%
)=R_{l}(r)Y_{l_{1}\cdots l_{D-2}}^{(l)}(\theta _{1},\theta _{2},\cdots
,\theta _{D-1}),
\end{equation}%
where $E_{nl},$ $M$ and $\mathbf{\nabla }_{D}^{2}$ denote the KG energy, the
mass and the $D$-dimensional Laplacian, respectively. In addition, $\mathbf{x%
}$ is a $D$-dimensional position vector. Let us decompose the radial wave
function $R_{l}(r)$ as follows:%
\begin{equation}
R_{l}(r)=r^{-(D-1)/2}u_{l}(r),
\end{equation}%
we, then, reduce Eq. (30) into the $D$-dimensional radial Schr\"{o}%
dinger-like equation with arbitrary orbital angular momentum number $l$ as

\begin{equation}
\frac{d^{2}u_{l}(r)}{dr^{2}}+\frac{1}{\hbar ^{2}c^{2}}\left\{ \left[
E_{nl}-V(r)\right] ^{2}-\left[ Mc^{2}+S(r)\right] ^{2}-\frac{l^{\prime
}(l^{\prime }+1)\hbar ^{2}c^{2}}{r^{2}}\right\} u_{l}(r)=0,
\end{equation}%
where we have set $l^{\prime }(l^{\prime }+1)=\left[ (\mathcal{M}-2)^{2}-1%
\right] /4$ and $\mathcal{M}=D+2l$ where $l=0,1,2,\cdots .$ Under the
equally mixed potentials $S(r)=\pm V(r),$ the KG turns into a Schr\"{o}%
dinger-like equation and thus the bound state solutions are very easily
obtained with the help of the well-known methods developed in the
non-relativistic quantum mechanics. We use the existing approximation for
the centrifugal potential term in the non-relativistic model [7,14] which is
valid only for $q=1$ value [49,55]:%
\begin{equation}
\widetilde{V}(r)=\frac{l^{\prime }(l^{\prime }+1)}{r^{2}}\approx 4\alpha
^{2}l^{\prime }(l^{\prime }+1)\frac{e^{-2\alpha r}}{\left( 1-qe^{-2\alpha
r}\right) ^{2}},\text{ }l^{\prime }=\left( \mathcal{M}-3\right) /2,
\end{equation}%
in the limit of small $\alpha $ and $l^{\prime }.$

\subsection{The Eckart-type model}

At first, let us rewrite Eq. (2) in a form to include the Hulth\'{e}n
potential,

\begin{equation}
V(r;q)=4V_{1}\frac{e^{-2\alpha r}}{\left( 1-qe^{-2\alpha r}\right) ^{2}}%
-V_{2}\frac{1}{1-qe^{-2\alpha r}}-V_{3}\frac{qe^{-2\alpha r}}{1-qe^{-2\alpha
r}},
\end{equation}%
and then follow the model used in Refs. [49,55,56] by inserting the above
equation and the approximate potential term (33) into (32), we obtain%
\begin{equation*}
\frac{d^{2}u_{l}(r)}{dr^{2}}+
\end{equation*}%
\begin{equation*}
\frac{1}{\hbar ^{2}c^{2}}\left\{ -\frac{\left[ 8\left( E_{nl}\pm
Mc^{2}\right) V_{1}+4\alpha ^{2}\hbar ^{2}c^{2}l^{\prime }(l^{\prime }+1)%
\right] e^{-2\alpha r}}{\left( 1-qe^{-2\alpha r}\right) ^{2}}+\frac{2\left(
E_{nl}\pm Mc^{2}\right) \left( V_{2}+qV_{3}e^{-2\alpha r}\right) }{\left(
1-qe^{-2\alpha r}\right) }\right\} u_{l}(r)
\end{equation*}%
\begin{equation}
=\frac{1}{\hbar ^{2}c^{2}}\left[ \left( Mc^{2}\right) ^{2}-E_{nl}^{2}\right]
u_{l}(r),\text{ }u_{l}(0)=0,
\end{equation}%
which is now amenable to the NU solution. We further use the following ans%
\"{a}tze in order to make the above differential equation more compact%
\begin{equation*}
z(r)=e^{-2\alpha r},\text{ }\varepsilon _{nl}=\frac{\sqrt{\left(
Mc^{2}\right) ^{2}-E_{nl}^{2}}}{Q},\text{ }\beta =\frac{8\left( E_{nl}\pm
Mc^{2}\right) V_{1}}{Q^{2}}+l^{\prime }(l^{\prime }+1),
\end{equation*}%
\begin{equation}
\text{ }\gamma =\frac{2\left( E_{nl}\pm Mc^{2}\right) V_{2}}{Q^{2}},\text{ }%
\lambda =\frac{2\left( E_{nl}\pm Mc^{2}\right) V_{3}}{Q^{2}},\text{ }%
Q=2\hbar c\alpha .
\end{equation}%
Notice that $\left\vert E_{nl}\right\vert \leq Mc^{2}.$ The KG equation can
then be reduced to%
\begin{equation*}
\left[ z(1-qz)\right] ^{2}\frac{d^{2}u_{l}(z)}{dz^{2}}+z(1-qz)^{2}\frac{%
du_{l}(z)}{dz}
\end{equation*}%
\begin{equation}
+\left\{ -q^{2}(\varepsilon _{nl}^{2}+\lambda )z^{2}+(2q\varepsilon
_{nl}^{2}+q\lambda -q\gamma -\beta )z-\left( \varepsilon _{nl}^{2}-\gamma
\right) \right\} u_{l}(z)=0,
\end{equation}%
where $r\in \lbrack 0,\infty )\rightarrow z\in \lbrack 0,1].$ Before
proceeding, the boundary conditions on the radial wave function $u_{l}(r)$
demand that $u_{l}(r\rightarrow \infty $ or $z\rightarrow 0)\rightarrow 0$
and $u_{l}(r=0$ or $z=1)$ is finite. Comparing Eq. (37) with Eq. (13), we
obtain values for the set of parameters given in section 2:%
\begin{equation*}
c_{1}=1,\text{ }c_{2}=c_{3}=q,\text{ c}_{4}=0,\text{ }c_{5}=-\frac{q}{2},%
\text{ }c_{6}=q^{2}\left( \varepsilon _{nl}^{2}+\lambda +\frac{1}{4}\right) ,
\end{equation*}%
\begin{equation*}
c_{7}=-q\left( 2\varepsilon _{nl}^{2}+\lambda -\gamma -\frac{\beta }{q}%
\right) ,\text{ }c_{8}=\varepsilon _{nl}^{2}-\gamma ,\text{ }c_{9}=\left( 
\frac{q}{2}\right) ^{2}\left( 1+\frac{4\beta }{q}\right) ,\text{ }c_{10}=2%
\sqrt{\varepsilon _{nl}^{2}-\gamma },
\end{equation*}%
\begin{equation*}
c_{11}=\sqrt{1+\frac{4\beta }{q}},\text{ }c_{12}=\sqrt{\varepsilon
_{nl}^{2}-\gamma },\text{ }c_{13}=\frac{1}{2}\left( 1+\sqrt{1+\frac{4\beta }{%
q}}\right) ,
\end{equation*}%
\begin{equation}
A=q^{2}\left( \varepsilon _{nl}^{2}+\lambda \right) ,\text{ }%
B=q(2\varepsilon _{nl}^{2}+\lambda -\gamma -\frac{\beta }{q}),\text{ }%
C=\varepsilon _{nl}^{2}-\gamma ,
\end{equation}%
and the energy equation via Eq. (23) as

\begin{equation}
\varepsilon _{nl}^{2}=\frac{\left( \gamma +\lambda \right) ^{2}}{4(n+\delta
)^{2}}+\frac{\left( n+\delta \right) ^{2}}{4}+\frac{\gamma -\lambda }{2},%
\text{ }n=0,1,2,\cdots ,
\end{equation}%
where $\delta =\frac{1}{2}\left( 1+\sqrt{1+\frac{4\beta }{q}}\right) .$
Making use of Eq. (36), the above equation turns to become

\begin{equation}
M^{2}c^{4}-E_{nl}^{2}=\left( \hbar c\alpha \right) ^{2}\left( n+w\right)
^{2}+\frac{\left( E_{nl}\pm Mc^{2}\right) ^{2}}{\left( 2\hbar c\alpha
\right) ^{2}}\frac{\left( V_{2}+V_{3}\right) ^{2}}{\left( n+w\right) ^{2}}%
+\left( E_{nl}\pm Mc^{2}\right) \left( V_{2}-V_{3}\right) ,
\end{equation}%
where $w=\frac{1}{2}\left( 1+\sqrt{1+\frac{4l^{\prime }(l^{\prime }+1)}{q}+%
\frac{8\left( E_{nl}\pm Mc^{2}\right) V_{1}}{q\left( \hbar c\alpha \right)
^{2}}}\right) .$ The energy $E_{nl}$ is defined implicitly by Eq. (40) which
is a rather complicated transcendental equation having many solutions for
given values of $n$ and $l.$ In the above equation, let us remark that it is
not difficult to conclude that bound-states appear in four energy solutions;
only two energy solutions are valid for the particle $E^{p}=E_{nl}^{+}$ and
the second one corresponds to the anti-particle energy $E^{a}=E_{nl}^{-}$ in
the Eckart-type field.

Referring to the general parametric model in section 2, we can also
calculate the corresponding wave functions. The explicit form of the weight
function becomes%
\begin{equation}
\rho (z)=z^{2p}(1-qz)^{2w-1},\text{ }p=\frac{1}{2}\left[ n+w-\frac{\left(
E_{nl}\pm Mc^{2}\right) \left( V_{2}+V_{3}\right) }{2\left( \hbar c\alpha
\right) ^{2}}\frac{1}{n+w}\right] ,\text{ }
\end{equation}%
which gives the following Jacobi polynomials:%
\begin{equation}
y_{n}(z)\rightarrow P_{n}^{\left( 2p,2w-1\right) }(1-2qz),
\end{equation}%
as\ a first part of the wave functions. The second part of the wave
functions can be found as%
\begin{equation}
\phi (z)\rightarrow z^{p}(1-qz)^{w}.
\end{equation}%
Hence, the unnormalized wave functions expressed in terms of the Jacobi
polynomials read%
\begin{equation}
u_{l}(z)=\mathcal{N}_{n}z^{p}(1-qz)^{w}P_{n}^{\left( 2p,2w-1\right) }(1-2qz),
\end{equation}%
and consequently the total radial part of the wave functions expressed in
terms of the hypergeometric functions are 
\begin{equation}
R_{l}(r)=\mathcal{N}_{n}r^{-(D-1)/2}\left( e^{-2\alpha r}\right)
^{p}(1-qe^{-2\alpha r})^{w}%
\begin{array}{c}
_{2}F_{1}%
\end{array}%
(-n,n+2\left( p+w\right) ;2p+1;qe^{-2\alpha r}),
\end{equation}%
where $\mathcal{N}_{n}$ is a constant related to the normalization. The
relationship between the Jacobi polynomials and the hypergeometric functions
is given by $P_{n}^{\left( a,b\right) }(1-2qx)=_{2}F_{1}(-n,n+a+b+1;a+1;x),$
where $_{2}F_{1}(\nu ,\mu ;\gamma ;x)=\frac{\Gamma (\gamma )}{\Gamma (\nu
)\Gamma (\mu )}\dsum\limits_{k=0}^{\infty }$ $\frac{\Gamma (\nu +k)\Gamma
(\mu +k)}{\Gamma (\gamma +k)}\frac{x^{k}}{k!}.$

Now, when taking $V_{2}=V_{3},$ the energy equation (40) satisfying $E_{nl}$
for the equally-mixed scalar and vector Eckart-type potentials becomes%
\begin{equation}
M^{2}c^{4}-E_{nl}^{2}=\left( \hbar c\alpha \right) ^{2}\left( n+w\right)
^{2}+\frac{\left( E_{nl}\pm Mc^{2}\right) ^{2}}{\left( \hbar c\alpha \right)
^{2}}\frac{V_{2}^{2}}{\left( n+w\right) ^{2}},
\end{equation}%
and the wave functions:%
\begin{equation}
u_{l}(z)=\mathcal{N}_{n}z^{\upsilon }(1-qz)^{w}P_{n}^{\left( 2\upsilon
,2w-1\right) }(1-2qz),\text{ }\upsilon =\frac{1}{2}\left[ n+w-\frac{\left(
E_{nl}\pm Mc^{2}\right) V_{2}}{\left( \hbar c\alpha \right) ^{2}}\frac{1}{n+w%
}\right] ,
\end{equation}%
or the total radial wave functions in (30) are 
\begin{equation}
R_{l}(r)=\mathcal{N}_{n}r^{-(D-1)/2}\left( e^{-2\alpha r}\right) ^{\upsilon
}(1-qe^{-2\alpha r})^{w}%
\begin{array}{c}
_{2}F_{1}%
\end{array}%
(-n,n+2\left( \upsilon +w\right) ;2\upsilon +1;qe^{-2\alpha r}),
\end{equation}%
where $\mathcal{N}_{n}$ is a normalization factor. The results given in Eqs.
(46) and (47) are consistent with those given in Eqs. (15) and (18) of Ref.
[45].

Also, in taking $q=1,$ $2\alpha \rightarrow \alpha ,$ $V_{1}=V_{2}=0$ and $%
V_{3}=V_{0},$ Eq. (34) turns to become the Hulth\'{e}n potential. Hence, we
find bound state solutions for equally-mixed scalar and vector $S(r)=V(r)$
Hulth\'{e}n potentials in the KG theory with any orbital angular momentum
quantum number $l$ and an arbitrary dimension $D,$%
\begin{equation}
\sqrt{M^{2}c^{4}-E_{nl}^{2}}=\frac{\left( \hbar c\alpha \right) \left( n+\nu
\right) }{2}-\frac{\left( Mc^{2}+E_{nl}\right) V_{0}}{\hbar c\alpha }\frac{1%
}{\left( n+\nu \right) },\text{ }\nu =\frac{D+2l-1}{2},
\end{equation}%
\begin{equation}
u_{l}(z)=\mathcal{N}_{n}\left( e^{-\alpha r}\right) ^{\varsigma
}(1-e^{-\alpha r})^{\nu }P_{n}^{\left( 2\varsigma ,2\nu -1\right) }(1-2z),%
\text{ }\varsigma =\frac{n+\nu }{2}-\frac{\left( Mc^{2}+E_{nl}\right) V_{0}}{%
\left( \hbar c\alpha \right) ^{2}}\frac{1}{n+\nu },\text{ }
\end{equation}%
and the Jacobi polynomial in the above equation can be expressed in terms of
the hypergeometric function: 
\begin{equation}
R_{l}(r)=\mathcal{N}_{n}r^{-(D-1)/2}\left( e^{-\alpha r}\right) ^{\varsigma
}(1-e^{-\alpha r})^{\nu }%
\begin{array}{c}
_{2}F_{1}%
\end{array}%
(-n,n+2\left( \varsigma +\nu \right) ;2\varsigma +1;e^{-\alpha r}),
\end{equation}%
where $\mathcal{N}_{n}$ is a constant related to the normalization. The
above results are identical to those found recently by Refs. [49,57].

In the non-relativistic limit, inserting the equally mixed Eckart-type
potentials (1) into the Schr\"{o}dinger equation gives%
\begin{equation}
\frac{d^{2}u_{l}(r)}{dr^{2}}+\left\{ \frac{2ME_{nl}}{\hbar ^{2}}-\frac{\left[
8MV_{1}+4\alpha ^{2}\hbar ^{2}l^{\prime }(l^{\prime }+1)\right] e^{-2\alpha
r}}{\hbar ^{2}\left( 1-qe^{-2\alpha r}\right) ^{2}}+\frac{2MV_{2}\left(
1+qe^{-2\alpha r}\right) }{\hbar ^{2}\left( 1-qe^{-2\alpha r}\right) }%
\right\} u_{l}(r)=0,
\end{equation}%
and further making use of the following definitions:%
\begin{equation}
\varepsilon _{nl}=\frac{\sqrt{-2ME_{nl}}}{T},\text{ \ }E_{nl}\leq 0,\text{ }%
\beta =\frac{8MV_{1}}{T^{2}}+l^{\prime }(l^{\prime }+1),\text{ }\gamma =%
\frac{2MV_{2}}{T^{2}},\text{ }T=2\hbar \alpha ,
\end{equation}%
lead us to obtain the set of parameters and energy equation given before in
Eqs. (38) and (39) with $\gamma =\lambda $. Incorporating the above equation
and using Eq. (39), we find the following energy eigenvalues:%
\begin{equation}
E_{nl}=-\frac{1}{2M}\left[ \hbar ^{2}\alpha ^{2}\left( n+w_{1}\right) ^{2}+%
\frac{M^{2}V_{2}^{2}}{\hbar ^{2}\alpha ^{2}}\frac{1}{\left( n+w_{1}\right)
^{2}}\right] ,\text{ }w_{1}=\frac{1}{2}\left( 1+\sqrt{(1+2l^{\prime })^{2}+%
\frac{8MV_{1}}{\hbar ^{2}\alpha ^{2}}}\right)
\end{equation}%
In addition, following the procedures indicated in Eqs. (41)-(45), we obtain
expressions for the radial wave functions:%
\begin{equation*}
R_{l}(r)=\mathcal{N}_{n}^{\prime }r^{-(D-1)/2}\left( e^{-2\alpha r}\right)
^{p_{1}}(1-e^{-2\alpha r})^{w_{1}}P_{n}^{\left( 2p_{1},2w_{1}-1\right)
}(1-2e^{-2\alpha r}),
\end{equation*}%
\begin{equation}
p_{1}=\frac{1}{2\hbar \alpha }\sqrt{-2M\left( E_{nl}+V_{2}\right) }=\frac{1}{%
2}\left[ n+w_{1}-\frac{MV_{2}}{\hbar ^{2}\alpha ^{2}}\frac{1}{n+w_{1}}\right]
.
\end{equation}

\subsection{The Rosen-Morse-type model}

Under the replacement of $q$ by $-q,$ the Eckart-type potential model given
in Eq. (1) will become the Rosen-Morse-type potential model given in Eq. (2)
of Ref. [39]:

\begin{equation}
V(r,q)=V_{1}\sec h_{q}^{2}\alpha r-V_{2}\tanh _{q}\alpha r,\text{ }%
V_{1},V_{2}>0,
\end{equation}%
or alternatively [26,58]%
\begin{equation}
V(r,q)=4V_{1}\frac{e^{-2\alpha r}}{\left( 1+qe^{-2\alpha r}\right) ^{2}}%
-V_{2}\frac{1-qe^{-2\alpha r}}{1+qe^{-2\alpha r}}.
\end{equation}%
We may rewrite the above equation in a form to include the Woods-Saxon
potential,%
\begin{equation}
V(r,q)=4V_{1}\frac{e^{-2\alpha r}}{\left( 1+qe^{-2\alpha r}\right) ^{2}}%
-V_{2}\frac{1}{1+qe^{-2\alpha r}}+V_{3}\frac{qe^{-2\alpha r}}{1+qe^{-2\alpha
r}}.
\end{equation}%
Using the following definitions 
\begin{equation*}
\varepsilon _{n,0}=\frac{\sqrt{\left( Mc^{2}\right) ^{2}-E_{n,0}^{2}}}{Q},%
\text{ }\widetilde{\beta }=\beta (l\rightarrow 0)=\frac{8V_{1}\left(
E_{n,0}\pm Mc^{2}\right) }{Q^{2}},\text{ }
\end{equation*}%
\begin{equation}
\widetilde{\gamma }=\gamma (l\rightarrow 0)=\frac{2\left( E_{n,0}\pm
Mc^{2}\right) V_{2}}{Q^{2}},\text{ }\widetilde{\lambda }=\lambda
(l\rightarrow 0)=\frac{2\left( E_{n,0}\pm Mc^{2}\right) V_{3}}{Q^{2}},
\end{equation}%
we write the $s$-wave KG equation with $S(r)=\pm V(r)$ for the potential
(58) as%
\begin{equation*}
\left[ z(1+qz)\right] ^{2}\frac{d^{2}u_{n}(z)}{dz^{2}}+z(1+qz)^{2}\frac{%
du_{n}(z)}{dz}
\end{equation*}%
\begin{equation}
+\left\{ -q^{2}\left( \varepsilon _{n,0}^{2}+\widetilde{\lambda }\right)
z^{2}+q\left( \widetilde{\gamma }-\widetilde{\lambda }-2\varepsilon
_{n,0}^{2}-\frac{\widetilde{\beta }}{q}\right) z-\left( \varepsilon
_{n,0}^{2}-\widetilde{\gamma }\right) \right\} u_{n}(z)=0.
\end{equation}%
Following same procedures used in the previous subsection, we obtain values
for the parameters given in section 2:%
\begin{equation*}
c_{1}=1,\text{ }c_{2}=c_{3}=-q,\text{ c}_{4}=0,\text{ }c_{5}=\frac{q}{2},%
\text{ }c_{6}=q^{2}\left( \varepsilon _{n,0}^{2}+\widetilde{\lambda }+\frac{1%
}{4}\right) ,
\end{equation*}%
\begin{equation*}
c_{7}=q\left( 2\varepsilon _{n,0}^{2}+\widetilde{\lambda }+\frac{\widetilde{%
\beta }}{q}-\widetilde{\gamma }\right) ,\text{ }c_{8}=\varepsilon _{n,0}^{2}-%
\widetilde{\gamma },\text{ }c_{9}=\left( \frac{q}{2}\right) ^{2}\left( 1-%
\frac{4\widetilde{\beta }}{q}\right) ,\text{ }c_{10}=2\sqrt{\varepsilon
_{n,0}^{2}-\widetilde{\gamma }},
\end{equation*}%
\begin{equation*}
c_{11}=-\sqrt{1-\frac{4\widetilde{\beta }}{q}},\text{ }c_{12}=\sqrt{%
\varepsilon _{n,0}^{2}-\widetilde{\gamma }},\text{ }c_{13}=\widetilde{\delta 
}=\frac{1}{2}\left( 1-\sqrt{1-\frac{4\widetilde{\beta }}{q}}\right) ,
\end{equation*}%
\begin{equation}
A=q^{2}\left( \varepsilon _{n,0}^{2}+\widetilde{\lambda }\right) ,\text{ }%
B=-q\left( 2\varepsilon _{n,0}^{2}+\frac{\widetilde{\beta }}{q}+\widetilde{%
\lambda }-\widetilde{\gamma }\right) ,\text{ }C=\varepsilon _{n,0}^{2}-%
\widetilde{\gamma },
\end{equation}%
and the energy equation

\begin{equation}
\varepsilon _{n,0}^{2}=\frac{\left( \widetilde{\gamma }+\widetilde{\lambda }%
\right) ^{2}}{4(n+\widetilde{\delta })^{2}}+\frac{\left( n+\widetilde{\delta 
}\right) ^{2}}{4}+\frac{\widetilde{\gamma }-\widetilde{\lambda }}{2}.
\end{equation}%
Inserting Eq. (59) in the above equation, we obtain energy equation
satisfying $E_{n,0},$

\begin{equation*}
M^{2}c^{4}-E_{n,0}^{2}=\left( \hbar c\alpha \right) ^{2}\left( n+\widetilde{w%
}\right) ^{2}+\frac{\left( E_{n,0}\pm Mc^{2}\right) ^{2}}{\left( 2\hbar
c\alpha \right) ^{2}}\frac{\left( V_{2}+V_{3}\right) ^{2}}{\left( n+%
\widetilde{w}\right) ^{2}}+\left( E_{n,0}\pm Mc^{2}\right) \left(
V_{2}-V_{3}\right) ,
\end{equation*}%
\begin{equation}
\widetilde{w}=\frac{1}{2}\left( 1-\sqrt{1-\frac{8\left( E_{n,0}\pm
Mc^{2}\right) V_{1}}{q\left( \hbar c\alpha \right) ^{2}}}\right) .
\end{equation}%
The corresponding unnormalized wave functions can be calculated as before,
the explicit form of the weight function becomes%
\begin{equation}
\rho (z)=z^{2\widetilde{p}}(1-qz)^{2\widetilde{w}-1},\text{ }\widetilde{p}=%
\frac{1}{2}\left[ n+\widetilde{w}-\frac{\left( E_{n,0}\pm Mc^{2}\right)
\left( V_{2}+V_{3}\right) }{2\left( \hbar c\alpha \right) ^{2}}\frac{1}{n+%
\widetilde{w}}\right] ,
\end{equation}%
which gives the Jacobi polynomials:%
\begin{equation}
y_{n}(z)\rightarrow P_{n}^{\left( 2\widetilde{p},2\widetilde{w}-1\right)
}(1+2qz),
\end{equation}%
as the first part of the wave function. The second part of the wave function
can be found as%
\begin{equation}
\phi (z)\rightarrow z^{\widetilde{p}}(1+qz)^{\widetilde{w}}.
\end{equation}%
The unnormalized wave function reads%
\begin{equation}
u_{n}(z)=\widetilde{\mathcal{N}}_{n}z^{\widetilde{p}}(1+qz)^{\widetilde{w}%
}P_{n}^{\left( 2\widetilde{p},2\widetilde{w}-1\right) }(1+2qz),
\end{equation}%
and thus the total radial part of the radial wave functions in (30) can be
expressed in terms of the hypergeometric functions as 
\begin{equation}
R_{n}(r)=\widetilde{\mathcal{N}}_{n}\left( e^{-2\alpha r}\right) ^{%
\widetilde{p}}(1+qe^{-2\alpha r})^{\widetilde{w}}%
\begin{array}{c}
_{2}F_{1}%
\end{array}%
(-n,n+2\left( \widetilde{p}+\widetilde{w}\right) ;2\widetilde{p}%
+1;-qe^{-2\alpha r}),
\end{equation}%
where $\widetilde{\mathcal{N}}_{n}$ is a normalization factor.

In taking $V_{2}=V_{3}$ in Eq. (63)$,$ we find the equation for the
potential in (56) satisfying $E_{n,0}$ in the $s$-wave KG theory,%
\begin{equation}
M^{2}c^{4}-E_{n,0}^{2}=\left( \hbar c\alpha \right) ^{2}\left( n+\widetilde{w%
}\right) ^{2}+\frac{\left( E_{n,0}\pm Mc^{2}\right) ^{2}}{\left( \hbar
c\alpha \right) ^{2}}\frac{V_{2}^{2}}{\left( n+\widetilde{w}\right) ^{2}},
\end{equation}%
and the wave functions are%
\begin{equation*}
u_{n}(r)=\widetilde{\mathcal{N}}_{n}\left( e^{-2\alpha r}\right) ^{%
\widetilde{p}_{1}}(1+qe^{-2\alpha r})^{\widetilde{w}}P_{n}^{\left( 2%
\widetilde{p}_{1},2\widetilde{w}-1\right) }(1+2qe^{-2\alpha r}),
\end{equation*}%
\begin{equation*}
=\widetilde{\mathcal{N}}_{n}\left( e^{-2\alpha r}\right) ^{\widetilde{p}%
_{1}}(1+qe^{-2\alpha r})^{\widetilde{w}}%
\begin{array}{c}
_{2}F_{1}%
\end{array}%
(-n,n+2\left( \widetilde{p}_{1}+\widetilde{w}\right) ;2\widetilde{p}%
_{1}+1;-qe^{-2\alpha r}),
\end{equation*}%
\begin{equation}
\widetilde{p}_{1}=\frac{1}{2}\left[ n+\widetilde{w}-\frac{\left( E_{n,0}\pm
Mc^{2}\right) V_{2}}{\left( \hbar c\alpha \right) ^{2}}\frac{1}{n+\widetilde{%
w}}\right] ,
\end{equation}%
where $\widetilde{\mathcal{N}}_{n}$ is a normalization constant. After
making appropriate change of the potential parameter $V_{1}\rightarrow
-V_{1} $ in Eq. (56), our results in Eqs. (69) and (70) become identical
with Eqs. (13) and (14) of Ref. [39].

Also, taking $q=1,$ $2\alpha \rightarrow \alpha ,$ $V_{1}=V_{2}=0$ and $%
V_{3}=-V_{0},$ Eq. (58) turns to become the Woods-Saxon potential. Hence, we
can find bound state solutions in the $s$-wave KG theory with equally-mixed
scalar and vector $S(r)=V(r)$ for Woods-Saxon potentials as%
\begin{equation}
\sqrt{M^{2}c^{4}-E_{n0}^{2}}=\hbar c\alpha \widetilde{p}_{2},\text{ }%
\widetilde{p}_{2}=\frac{n}{2}+\frac{\left( Mc^{2}+E_{n0}\right) V_{0}}{%
\left( \hbar c\alpha \right) ^{2}}\frac{1}{n},
\end{equation}%
and wave functions:%
\begin{equation}
u_{n}(r)=\mathcal{N}_{n}\left( e^{-\alpha r}\right) ^{\widetilde{p}%
_{2}}P_{n}^{\left( 2\widetilde{p}_{2},-1\right) }(1+2e^{-\alpha r}),
\end{equation}%
or alternatively, it can be expressed in terms of the hypergeometric
function as 
\begin{equation}
R_{n}(r)=\mathcal{N}_{n}r^{-1}\left( e^{-\alpha r}\right) ^{\widetilde{p}%
_{2}}%
\begin{array}{c}
_{2}F_{1}%
\end{array}%
(-n,n+2\widetilde{p}_{2};2\widetilde{p}_{2}+1;e^{-\alpha r}),
\end{equation}%
where $\mathcal{N}_{n}$ is a constant related to the normalization. Under
appropriate parameter replacements, we obtain the non-relativistic limit of
the energy eigenvalues and eigenfunctions of the above two equations are%
\begin{equation}
E_{n,0}=-\frac{1}{2M}\left[ \frac{n\hbar \alpha }{2}+\frac{2MV_{0}}{\hbar
\alpha }\frac{1}{n}\right] ,\text{ }n\neq 0,
\end{equation}%
and%
\begin{equation}
u_{n}(r)=\mathcal{N}_{n}\left( e^{-\alpha r}\right) ^{\widetilde{p}_{3}}%
\begin{array}{c}
_{2}F_{1}%
\end{array}%
(-n,n+2\widetilde{p}_{2};2\widetilde{p}_{2}+1;e^{-\alpha r}),\text{ }%
\widetilde{p}_{3}=\frac{n}{2}+\frac{2MV_{0}}{\left( \hbar c\alpha \right)
^{2}}\frac{1}{n},
\end{equation}%
respectively, which is the solution of the Schr\"{o}dinger equation for the
potential $\Sigma (r)=V(r)+S(r)=2V(r).$The above results are identical to
those found before by Ref. [6].

\section{Discussions}

In this section, at first, we choose appropriate parameters in the
Eckart-type potential model to construct the Eckart potential, Rosen-Morse
well and their PT-symmetric versions, and then discuss their energy
equations in the framework of KG theory with equally mixed potentials.

\subsection{Eckart potential}

Taking $q=1,$ the potential (1) turns to the standard Eckart potential [27]%
\begin{equation}
V(r)=V_{1}\cos ech^{2}\alpha r-V_{2}\coth \alpha r,\text{ }V_{1},V_{2}>0.
\end{equation}%
In natural units ($\hbar =c=1$), we can obtain the energy equation (46) for
the Eckart potential in the three-dimensional spinless KG theory as

\begin{equation*}
M^{2}-E_{nl}^{2}=\alpha ^{2}\left( n+w^{\prime }\right) ^{2}+\frac{\left(
E_{nl}\pm M\right) ^{2}}{\alpha ^{2}}\frac{V_{2}^{2}}{\left( n+w^{\prime
}\right) ^{2}},
\end{equation*}%
\begin{equation}
w^{\prime }=w(q\rightarrow 1)=\frac{1}{2}\left( 1+\sqrt{(2l^{\prime }+1)^{2}+%
\frac{8\left( E_{nl}\pm M\right) V_{1}}{\alpha ^{2}}}\right) ,
\end{equation}%
which is identical with those given in Eq. (22) of Ref. [39] under the
equally-mixed potential restriction given by $S(r)=\pm V(r).$ The
unnormalized wave function corresponding to the energy levels is%
\begin{equation}
R_{l}(r)=\mathcal{N}_{nl}^{\prime }r^{-(D-1)/2}\left( e^{-2\alpha r}\right)
^{\upsilon }(1-e^{-2\alpha r})^{w^{\prime }}P_{n}^{\left( 2\upsilon
,2w^{\prime }-1\right) }(1-2e^{-2\alpha r}),
\end{equation}%
where $\mathcal{N}_{nl}^{\prime }$ is a normalization factor.

(i) For $s$-wave case, the centrifugal term $\frac{(D+2l-1)(D+2l-3)}{4r^{2}}%
=0$ and consequently the approximation term $(D+2l-1)(D+2l-3)\alpha ^{2}%
\frac{e^{-2\alpha r}}{\left( 1-e^{-2\alpha r}\right) ^{2}}=0,$ too. Thus,
the energy eigenvalues take the following simple form 
\begin{equation}
M^{2}-E_{n,0}^{2}=\alpha ^{2}\left( n+w_{1}\right) ^{2}+\frac{%
V_{2}^{2}\left( E_{n,0}\pm M\right) ^{2}}{\alpha ^{2}\left( n+w_{1}\right)
^{2}},\text{ }w_{1}=\frac{1}{2}\left( 1+\sqrt{1+\frac{8\left( E_{n,0}\pm
M\right) V_{1}}{\alpha ^{2}}}\right) .
\end{equation}%
(ii) In the non-relativistic approximation of the KG energy equation
(potential energies small compared to $Mc^{2}$ and $E\simeq Mc^{2}$) Eq.
(32) reduces into the form [59]%
\begin{equation}
-\frac{\hbar ^{2}}{2M}\frac{d^{2}u_{l}(r)}{dr^{2}}+\left\{ V(r)+S(r)-\frac{%
l^{\prime }(l^{\prime }+1)\hbar ^{2}}{r^{2}}\right\} u_{l}(r)=\left(
E-Mc^{2}\right) u_{l}(r).
\end{equation}%
When $V(r)=S(r),$ the energy eigenvalues obtained from Eq. (80) reduces to
those energy eigenvalues obtained from the solution of the Schr\"{o}dinger
equation for the sum potential $\Sigma (r)=2V(r).$ In other words, the
non-relativistic limit is the Schr\"{o}dinger-like equation for the
potential $\ 8V_{1}\frac{e^{-2\alpha r}}{\left( 1-e^{-2\alpha r}\right) ^{2}}%
-2V_{2}\frac{1+e^{-2\alpha r}}{1-e^{-2\alpha r}}.$ This can be achieved by
making the parameter replacements $M+E_{R}\rightarrow 2M$ and $%
E_{R}-M\rightarrow E_{NR},$ so the non-relativistic limit of our results in
Eq. (46) reduces to%
\begin{equation}
E_{NR}=-\frac{1}{2M}\left[ \alpha ^{2}\left( n+w_{2}\right) ^{2}+\frac{%
2M^{2}V_{2}^{2}}{\alpha ^{2}\left( n+w_{2}\right) ^{2}}\right] ,
\end{equation}%
and the corresponding wave functions in (48) become%
\begin{equation*}
R_{l}(r)=\mathcal{N}_{nl}^{\prime }r^{-(D-1)/2}\left( e^{-2\alpha r}\right)
^{\upsilon _{2}}(1-e^{-2\alpha r})^{w_{2}}P_{n}^{\left( 2\upsilon
_{2},2w_{2}-1\right) }(1-2e^{-2\alpha r}),
\end{equation*}%
\begin{equation}
\upsilon _{2}=\frac{1}{2}\left[ n+w_{2}-\frac{2MV_{2}}{\alpha ^{2}}\frac{1}{%
n+w_{2}}\right] ,\text{ }w_{2}=\frac{1}{2}\left( 1+\sqrt{(1+2l^{\prime
})^{2}+\frac{16MV_{1}}{\alpha ^{2}}}\right) .
\end{equation}%
The above two equations are identical with the NU solution of the Schr\"{o}%
dinger equation for a potential $V(r)$ (cf. Eqs. (54) and (55))$.$

\subsection{PT-symmetric Trigonometric Rosen-Morse (tRM) potential}

When we make the transformations of parameters as $\alpha \rightarrow
i\alpha ,$ $V_{2}\rightarrow -iV_{2},$ and $V_{1}\rightarrow -V_{1},$ and
using the relation between the trigonometric and the hyperbolic functions $%
\sin \left( i\alpha x\right) =i\sinh \left( \alpha x\right) ,$ the potential
(1) turns to become the PT-symmetric tRM potential [60]:%
\begin{equation}
V(x)=V_{1}\csc ^{2}\alpha x-V_{2}\cot \alpha x,\text{ }\func{Re}(V_{1})>0,%
\text{ }\alpha =\frac{\pi }{2d},\text{ }x=(0,d],
\end{equation}%
where $V_{1}=a(a+1)$ and $V_{2}=2b.$ This potential is displayed in Figure 1
which is nearly linear in $\pi /3<\alpha x<2\pi /3,$ Coulombic in $\pi
/90<\alpha x<\pi /30$ and infinite walls at $0$ and $\pi .$ So it might be a
prime candidate for an effective QCD potential. For a potential $V(x),$ when
one makes the transformation of $x\rightarrow -x$ and $i\rightarrow -i,$ if
the relation $V(-x)=V^{\ast }(x)$ exists, the potential $V(x)$ is said to be
PT-symmetric, where $P$ denotes parity operator (space reflection) and $T$
denotes time reversal [6,61]. Our point here is that $V(x)$ interpolates
between the Coulomb-and the infinite wall potential [62] going through an
intermediary region of linear-$x$-and harmonic-oscillator $x^{2}$ dependences%
$.$ To see this it is quite instructive to expand the potential in a Taylor
series which for appropriately small $x$ takes the form of a Coulomb-like
potential with a centrifugal-barrier like term, provided by the $\csc
^{2}\alpha x$ part [63],%
\begin{equation}
V(x)\approx -\frac{V_{2}}{\alpha x}+\frac{V_{1}}{\left( \alpha x\right) ^{2}}%
,\text{ }\alpha x\ll 1.
\end{equation}%
For $\alpha x<\pi /2$ we can then take the potential (84) plus a linear like
perturbation%
\begin{equation}
\Delta V(x)=V_{1}/3+V_{2}x/3,
\end{equation}%
as an approximation of tRM potential. The potential (83) obviously evolves
to an infinite wall as $\alpha x$ approaches the limits of the definition
interval $0<\alpha x<\pi ,$ due to the behavior of the $\cot \alpha x$ and $%
\csc \alpha x$ for $V_{1}>0.$ The potential is essential for the QCD
quark-gluon dynamics where the one gluon exchange gives rise to an effective
Coulomb-like potential, while the self gluon interactions produce a linear
potential as established by lattice QCD calculations of hadron properties
(Cornell potential) [64]. Finally, the infinite wall piece of the tRM
potential provides the regime suited for the asymptotical freedom of the
quarks. Now, making the corresponding parameter replacements in Eq. (46), we
end up with real energy equation for the above PT-symmetric version of the
Eckart-type potential\c{s} in the KG equation with equally mixed potentials,%
\begin{equation}
\left( Mc^{2}\right) ^{2}-E_{nl}^{2}=\frac{\left( E_{nl}\pm Mc^{2}\right)
^{2}}{\left( \hbar c\alpha \right) ^{2}}\frac{V_{2}^{2}}{\left( n+w\right)
^{2}}-\left( \hbar c\alpha \right) ^{2}\left( n+w\right) ^{2},
\end{equation}%
and the radial wave functions build up as 
\begin{equation*}
R_{l}(x)=\mathcal{N}_{n}x^{-(D-1)/2}\left( e^{+i2\alpha x}\right) ^{\upsilon
}(1-e^{+i2\alpha x})^{w}%
\begin{array}{c}
_{2}F_{1}%
\end{array}%
(-n,n+2\left( \upsilon +w\right) ;2\upsilon +1;e^{+i2\alpha x}),
\end{equation*}%
\begin{equation}
\upsilon =\frac{1}{2}\left[ n+w+i\frac{\left( E_{nl}\pm Mc^{2}\right) V_{2}}{%
\left( \hbar c\alpha \right) ^{2}}\frac{1}{n+w}\right] ,\text{ }w=\frac{1}{2}%
\left( 1+\sqrt{(1+2l^{\prime })^{2}+\frac{8\left( E_{nl}\pm Mc^{2}\right)
V_{1}}{\left( \hbar c\alpha \right) ^{2}}}\right)
\end{equation}

\subsection{Standard Rosen-Morse well}

Taking $q=1,$ $V_{1}\rightarrow -V_{1}$ ($\widetilde{\beta }\rightarrow -%
\widetilde{\beta }$) and $V_{2}\rightarrow -V_{2}$ ($\widetilde{\gamma }%
\rightarrow -\widetilde{\gamma }$)$,$ the potential (56) turns to the
standard Rosen-Morse well [26,58]%
\begin{equation}
V(r)=-V_{1}\sec h^{2}\alpha r+V_{2}\tanh \alpha r,\text{ }V_{1},V_{2}>0.
\end{equation}%
This potential is useful in discussing polyatomic molecular vibrational
energies. An example of its application to the vibrational states of $NH_{3}$
was given by Rosen and Morse in [26,58]. Making the corresponding parameter
replacements in Eq. (69), we obtain the energy equation for the Rosen-Morse
well in the $s$-wave KG theory with equally mixed potentials,%
\begin{equation}
M^{2}-E_{n0}^{2}=\alpha ^{2}\left( n+\widetilde{\delta }_{1}\right) ^{2}+%
\frac{\left( E_{n0}\pm M\right) ^{2}}{\alpha ^{2}}\frac{V_{2}^{2}}{\left( n+%
\widetilde{\delta }_{1}\right) ^{2}},\text{ }\widetilde{\delta }_{1}=\frac{1%
}{2}\left( 1-\sqrt{1+\frac{8\left( E_{n0}\pm M\right) V_{1}}{\alpha ^{2}}}%
\right) .
\end{equation}%
The unnormalized wave function corresponding to the energy levels is%
\begin{equation*}
u_{n}(r)=\widetilde{\mathcal{N}}_{n}^{\prime }\left( e^{-2\alpha r}\right) ^{%
\widetilde{\eta }_{1}}(1+e^{-2\alpha r})^{\widetilde{\delta }%
_{1}}P_{n}^{\left( 2\widetilde{\eta }_{1},2\widetilde{\delta }_{1}-1\right)
}(1+2e^{-2\alpha r}),
\end{equation*}%
\begin{equation}
\widetilde{\eta }_{1}=\frac{1}{2}\left[ n+\widetilde{\delta }_{1}+\frac{%
\left( E_{n,0}\pm M\right) V_{2}}{\alpha ^{2}}\frac{1}{n+\widetilde{\delta }%
_{1}}\right] ,
\end{equation}%
where $\widetilde{\mathcal{N}}_{n}^{\prime }$ is a normalization constant.
The results given in Eqs. (89) and (90) are consistent with those given in
Eqs. (19) and (20) of Ref. [39], respectively. The $s$-wave energy states of
the KG equation for the Rosen-Morse potential are calculated for a set of
selected values parameters in Table 1.

When $V_{0}=S_{0},$ the non-relativistic limit is the solution of the Schr%
\"{o}dinger equation for the potential $\ -8V_{1}\frac{e^{-2\alpha r}}{%
\left( 1+e^{-2\alpha r}\right) ^{2}}+2V_{2}\frac{1-e^{-2\alpha r}}{%
1+e^{-2\alpha r}}.$ In the non-relativistic limits, the energy spectrum is%
\begin{equation}
E_{NR}=-\frac{1}{2M}\left[ \alpha ^{2}\left( n+\widetilde{\delta }%
_{2}\right) ^{2}+\frac{4M^{2}V_{2}^{2}}{\alpha ^{2}\left( n+\widetilde{%
\delta }_{2}\right) ^{2}}\right] ,\text{ }\widetilde{\delta }_{2}=\frac{1}{2}%
\left( 1-\sqrt{1+\frac{16MV_{1}}{\alpha ^{2}}}\right) ,
\end{equation}%
and the wave functions are%
\begin{equation}
R_{l}(r)=\widetilde{\mathcal{N}}_{n}^{\prime }{}^{\prime }\left( e^{-2\alpha
r}\right) ^{\widetilde{\eta }_{2}}(1+e^{-2\alpha r})^{\widetilde{\delta }%
_{2}}P_{n}^{\left( 2\widetilde{\eta }_{2},2\widetilde{\delta }_{2}-1\right)
}(1+2e^{-2\alpha r}),\text{ }\widetilde{\eta }_{2}=\frac{1}{2}\left[ n+%
\widetilde{\delta }_{2}+\frac{2MV_{2}}{\alpha ^{2}}\frac{1}{n+\widetilde{%
\delta }_{2}}\right] .
\end{equation}

\section{Conclusions}

We have used a parametric generalization model derived from the NU to obtain
the analytic bound state solutions of the KG equation with any orbital
angular momentum quantum number $l$ for equally mixed scalar and vector
Eckart-type potentials. These calculations include energy equation and the
unnormalized wave functions being expressed in terms of the Jacobi
polynomials or the hypergeometric functions. Furthermore, making appropriate
changes in the Eckart-type potential parameters, one can generate new bound
state solutions for various types of the well-known molecular potentials
like the Rosen-Morse well [26], the Eckart potential, the Hulth\'{e}n
potential [9], the Woods-Saxon potential [5] and the Manning-Rosen potential
[22] and others. It is also noted that under the PT-symmetry, the
exponential potentials can be transformed into the trigonometric potentials
with real bound state solutions. The KG equation with equally mixed scalar
and vector Rosen-Morse-type potentials can be solved exactly for $s$-wave
bound states. In the relativistic model, the energy equations for these
potentials are complicated transcendental equations [26]. The
non-relativistic limits are obtained with a proper replacements of
parameters and/or by solving the original Schr\"{o}dinger equation. The
relativistic and non-relativistic results are identical with those ones
obtained in the literature through various methods.

\acknowledgments The partial support provided by the Near East University is
highly acknowledged.

\newpage

{\normalsize 
}

\bigskip

\bigskip \newpage

\bigskip {\normalsize 
}

\baselineskip= 2\baselineskip

\FRAME{ftbpFO}{0.0277in}{0.0277in}{0pt}{\Qct{Plot of the tRM potential [see
Eq. (83)] for a set of parameters $a=0.5$ and $b=17.0$.}}{}{Figure 4}{} 
\begin{table}[tbp]
\caption{The $s$-wave energy spectrum of the equally mixed scalar and vector
Rosen-Morse-type potentials.}%
\begin{tabular}{l}
\begin{tabular}{llllllllll}
\tableline$n$ & $\alpha $ & $q$ & $V_{1}$ & $V_{2}$ & $M$ & $E_{1}$%
\tablenotetext[1]{Same as in Ref. [39].} & $E_{2}$ & $E_{3}$ & $E_{4}$ \\ 
\tableline$1$ & $1$ & $1$ & $1$ & $-1$ & $4$ & $1.8137$\tablenotemark[1] & $%
-1.9140$\tablenotemark[1] & $-3.3923$\tablenotemark[1] & $-3.9088$%
\tablenotemark[1] \\ 
$2$ &  &  &  &  &  & $-2.2117$ & $-3.6791$ & $-$ & $-$ \\ 
$3$ &  &  &  &  &  & $-0.6606$ & $-3.3105$ & $-$ & $-$ \\ 
$4$ &  &  &  &  &  & $0.8879$ & $-2.7697$ & $-$ & $-$ \\ 
$5$ &  &  &  &  &  & $1.8766$ & $-1.9765$ & $-$ & $-$ \\ 
$1$ & $1$ & $1$ & $2$ & $-2$ & $5$ & $0.9989$ & $-3.7763$ & $-4.7275$ & $%
-4.9351$ \\ 
$2$ &  &  &  &  &  & $-4.1746$ & $-4.7795$ & $-$ & $-$ \\ 
$3$ &  &  &  &  &  & $-3.3814$ & $-4.5376$ & $-$ & $-$ \\ 
$4$ &  &  &  &  &  & $-2.3989$ & $-4.2008$ & $-$ & $-$ \\ 
$5$ &  &  &  &  &  & $-1.3083$ & $-3.7529$ & $-$ & $-$%
\end{tabular}
\\ 
\begin{tabular}{llllllllll}
$1$ & $0.5$ & $1$ & $1$ & $-1$ & $4$ & $1.9558$ & $-3.5288$ & $-3.8460$ & $%
-3.9773$ \\ 
$2$ &  &  &  &  &  & $1.9608$ & $-2.5367$ & $-3.5326$ & $-3.9216$ \\ 
$3$ &  &  &  &  &  & $1.2294$ & $-0.5126$ & $-3.0732$ & $-3.8358$ \\ 
$4$ &  &  &  &  &  & $-2.4823$ & $-3.7191$ & $-$ & $-$ \\ 
$5$ &  &  &  &  &  & $-1.7822$ & $-3.5695$ & $-$ & $-$ \\ 
$1$ & $1$ & $0.5$ & $1$ & $-1$ & $4$ & $1.5783$ & $-3.2245$ & $-3.6502$ & $%
-3.9258$ \\ 
$2$ &  &  &  &  &  & $1.9995$ & $-1.5367$ & $-2.9520$ & $-3.7496$ \\ 
$3$ &  &  &  &  &  & $-1.9529$ & $-3.4736$ & $-$ & $-$ \\ 
$4$ &  &  &  &  &  & $-0.7335$ & $-3.0839$ & $-$ & $-$ \\ 
$5$ &  &  &  &  &  & $0.5489$ & $-2.5528$ & $-$ & $-$ \\ 
\tableline &  &  &  &  &  &  &  &  & 
\end{tabular}%
\end{tabular}%
\end{table}

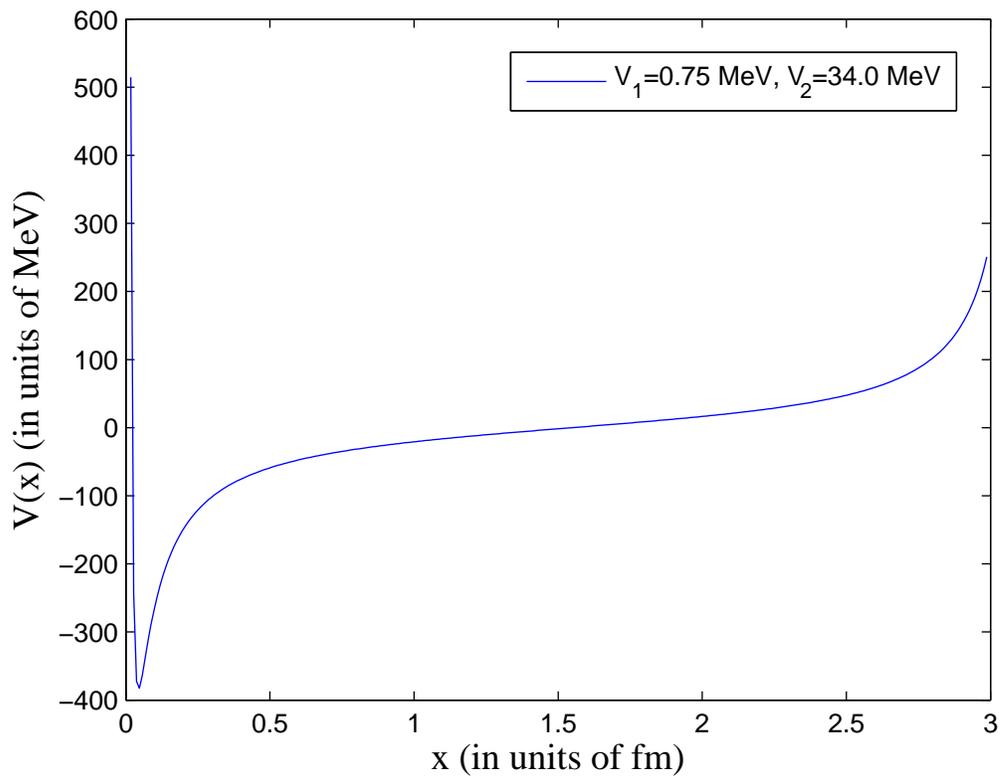
\end{document}